\documentclass[12pt, preprint]{aastex}
\usepackage{natbib}
\usepackage{graphicx}
\usepackage{subfigure}
\usepackage{amsmath,amssymb}
\begin{document}

\title{Elongation of Flare Ribbons}
%
%
%
\author{Jiong Qiu$^1$, Dana W. Longcope$^1$, Paul A. Cassak$^2$, Eric R. Priest$^3$}
\affil{1. Department of Physics, Montana State University, Bozeman MT USA\\
2. Department of Physics and Astronomy, West Virginia University, Morgantown WV, USA\\
3. School of Mathematics and Statistics, University of St. Andrews, Fife KY16 9SS, Scotland, UK}

\begin{abstract}
We present an analysis of the apparent elongation motion of flare ribbons along the polarity inversion line (PIL) as well as
the shear of flare loops in several two-ribbon flares. Flare ribbons and loops 
spread along the PIL at a speed ranging from a few to a hundred km s$^{-1}$. 
The shear measured from conjugate foot-points is consistent with
the measurement from flare loops, and both show the decrease of shear
toward a potential field as a flare evolves and ribbons and loops spread along the PIL.
Flares exhibiting fast bi-directional elongation appear to have a strong shear, which
may indicate a large magnetic guide field relative to the reconnection field in the coronal
current sheet. We discuss how  the analysis of ribbon motion could help infer properties in the 
corona where reconnection takes place.
\end{abstract}

\section{INTRODUCTION}
Two-ribbon flares have been used as textbook examples demonstrating the standard
flare reconnection configuration. The standard model is two-dimensional (2d), 
which would imply simultaneous reconnection everywhere along the entire current sheet. 
However, this does not occur in Nature, and all two-ribbon flares exhibit properties deviating 
from the 2d assumption. Even the most 2d-like two-ribbon flare arcade consists of 
discrete fine loops \citep{Aschwanden2001}, each of them formed by a reconnection
event. In this sense, reconnection is fragmented in space. The question is: do multiple 
reconnection events take place sporadically at several locations along the current sheet 
as a result of a global instability, or do adjacent reconnection events trigger one another in an organized manner? 
Observationally, many, though not all, flares exhibit localized reconnection events that
are globally organized. This is characterized by the apparent elongation motion of 
flare ribbons in the lower atmosphere along the magnetic polarity inversion line (PIL) 
-- the ``zipper" motion -- followed by an ordered spreading of flare loops anchored at the ribbons.
These observations reflect a manifestation of energy release and the formation of flare loops
by reconnection events in the corona successively along the PIL.

Spreading of flare ribbons or flare loops has been observed for several decades, and a 
vocabulary has been developed to describe this phenomenon. \citet{Vorpahl1976}
reported ``sequential brightening" of flare soft X-ray emission along the axis of a flare arcade, 
at an apparent speed of 180 -- 280 km s$^{-1}$. \citet{Kawaguchi1982} and \citet{Kitahara1990} 
reported ``progressive brightenings" of short duration H$\alpha$ emission along two flare ribbons, 
at a speed ranging from 100 to 500 km s$^{-1}$. These authors interpreted the apparent along-the-ribbon motion 
as due to propagating magnetosonic waves, although the estimated wave speed is a few times greater than 
the observed ribbon spreading speed. 

Subsequent observers have reported a large number of flares with sequential brightenings of flare 
loops along the magnetic PIL. Motivated by the dawn-dusk asymmetry of 
magnetospheric substorms, \citet{Isobe2002} examined ``successive formation" of soft X-ray loops along the arcade axis
observed by the {\it Solar X-ray Telescope} \citep[{\it SXT};][]{Tsuneta1991}. They found 21 events of such characters 
with a speed of 3 -- 30 km s$^{-1}$, substantially lower than those found in earlier case studies. They 
also found that 15 out of these 21 arcade events spread along the same 
direction as the reconnection electric field $\vec{E}$ in the corona, and the rest of them spread in the 
opposite direction. In a selective sample study, \citet{Tripathi2006} measured ``propagation" of flare EUV loops observed 
by the {\it Extreme ultraviole Imager Telescope} ({\em EIT}) at 195 \AA\ in 17 events which were accompanied by 
erupting filaments. In these events, the measured propagation 
speed ranges from 13 -- 111 km s$^{-1}$, yet mostly below 50 km s$^{-1}$. They also found that the propagation speed of flare loops is larger in some events where filament motion is relatively fast. In their sample, 15 out of the 17 events exhibit uni-directional propagation as well as asymmetric filament eruption (e.g., eruption from one end), and the other 2 exhibit bi-directional propagation and symmetric filament eruption (e.g., eruption from the middle).
\citet{Li2009} also studied ``propagation" of flare loops observed by the
{\it Transition Region and Coronal Explorer} \citep[{\it TRACE}; ][]{Handy1999} at 171 \AA\ from 1998 -- 2005, providing
a much larger and unbiased sample that includes 151 M-class and 39 X-class flares. The measured speed ranges
from 3 to 39 km s$^{-1}$, in the same range as \citet{Isobe2002}. \citet{Li2009} reported that about half of these 
events exhibit uni-directional propagation, and the other half exhibit bi-directional propagation.

Flare ribbons or kernels in the lower atmosphere outline the feet of flare loops or arcades, and are often brightened
impulsively in optical and UV wavelengths before flare loops are visible in soft X-ray and EUV wavelengths. 
They are therefore instantaneous signatures of energy release by reconnection. Propagation of ribbon fronts or kernels
along the polarity inversion line has been studied with much improved instrument tempo-spatial resolutions 
\citep{Fletcher2004, Lee2008, Qiu2009, Qiu2010, Liu2010, Cheng2012}. Most of these are case studies, and the 
measured apparent motion speed, ranging between 15 to 70 km$^{-1}$, is generally lower than those reported earlier by 
\citet{Kawaguchi1982} and \citet{Kitahara1990}, but higher than the measurements of loop propagation \citep{Isobe2002, Li2009}.

Finally,  apparent motion is also found in thick-target hard X-ray foot-point sources observed by the {\it Hard X-ray Telescope}
\citep[{\it HXT}; ][]{Kosugi1991} and the {\it Reuven Ramaty High Energy Solar Spectroscopic Imager} \citep[{\it RHESSI}; ][]{Lin2002}. 
\citet{Bogachev2005} examined flares observed with {\it HXT} at M2 band (photon energy 33 -- 53 keV) from 1991 through 2001, finding 
that out of 31 events that exhibit hard X-ray source motion, 11/8 events show conjugate sources moving in the same/opposite direction 
along the magnetic polarity inversion line, so-called ``parallel or antiparallel" motion. 
Using {\it RHESSI} observations, \citet{Krucker2003} observed one hard X-ray foot-point 
moving along the flare ribbon at a speed of 50~km s$^{-1}$. \citet{Grigis2005} found the apparent motion of a pair of 
conjugate foot-point sources with a mean speed of 50 - 150 km s$^{-1}$, with both sources moving in the same direction.
In a more comprehensive sample study, \citet{Yang2009} examined 27 {\it RHESSI} M-X class flares that exhibit motion 
of hard X-ray foot-points, finding that ``parallel/antiparallel motion" is present in most of these events during the 
impulsive phase of the flare (defined by the rise of the GOES 1-8 \AA\ soft X-ray emission), which is sometimes followed 
by the ``perpendicular motion" of the sources taking place more frequently in the gradual phase (the decay of the GOES 
soft X-ray emission). A two-step foot-point motion pattern was also noted by \citet{Qiu2009, Qiu2010, Cheng2012}. 
Specifically \citet{Qiu2010} and \citet{Cheng2012} studied two X-class flares, and showed that the newly brightened flare 
ribbon fronts observed in UV 1600 \AA\ by {\it TRACE} and the hard X-ray foot-points observed by {\it HXT} or {\it RHESSI} exhibit mostly 
consistent motion patterns in both wavelengths, first along the ribbon at a speed between 20 - 90 km s$^{-1}$, and 
then perpendicular to the ribbon. However, \citet{Inglis2013} found an X-class flare showing more complicated motion of the HXR foot-points,
such as a reversal of motion along the UV ribbon (note that such a reversal of motion was reported in \citet{Cheng2012} as well), 
and a mismatch between the brightest UV emissions and the HXR source in the early stage of the flare. It is likely that UV emission is produced
when the lower atmosphere is heated by either particle precipitation or thermal conduction \citep{Coyner2009}, both resulting from 
reconnection energy release; on the other hand, hard X-ray sources tend to be mapped to locations where energy flux is strong 
\citep{Temmer2007, Liu2007}. In this sense, morphology of flare UV emissions may reflect a more complete mapping of 
reconnection energy release in the lower atmosphere.

Without directly measuring the motion speed or directionality, some other studies report a ``shear motion" of conjugate flare foot-points.
Here ``shear" is defined as the angle made by the line connecting the conjugate foot-points with the line perpendicular
to the magnetic polarity inversion line of the photospheric magnetogram. \citet{Su2007} found that 87\% of selected 50 M-X class two-ribbon flares 
observed by {\it TRACE} exhibit the trend that the foot-point shear is strong at the onset of the flare, and
decreases as the flare evolves \citep[also see ][]{Yang2009}. Note that this shear measurement could arise from 
different motion patterns. It could be caused by conjugate foot-points moving in opposite directions along 
the ribbons, either approaching each other (shear decrease) or receding from each other (shear increase). This is different 
from the reported ``zipper" effect referring to the loops (and the conjugate feet of the loops) propagating in the same direction. 
The shear change can be also produced by conjugate foot-points moving in the same direction (i.e., zipper motion) but at 
different speeds along the PIL. 

All these observations of flare loop or ribbon motion along the ribbon direction indicate
the 3-dimensional nature of magnetic reconnection in solar flares. In a 2d framework, the along-the-ribbon 
motion is not present, and reconnection in the corona is characterized by a macroscopic electric field along the current sheet, 
assumed to be in the direction of the PIL, $\vec{E} = - \vec{v}_{in} \times \vec{B}_{in} \approx \vec{v}_{\perp}\times\vec{B_n}$ 
\citep{Forbes2000}. Here $\vec{v}_{in}$ and $\vec{B}_{in}$ are the inflow velocity and magnetic field vectors at the boundary of 
the reconnection diffusion region in the corona. The perpendicular motion of the ribbon, also referred as the ``separation motion", is given by $\vec{v}_{\perp}$, and $\vec{B}_n$ is the normal 
component of the 
magnetic field in the chromosphere where the ribbon is formed, pointing either inward or outward from the solar surface. 
This perpendicular motion has been reported in most flare studies cited above, and its speed has been measured in 
flares to range from a few to a few tens km s$^{-1}$. With some assumptions or approximations made about the coronal inflow magnetic 
field $B_{in}$ and plasma density $\rho$ around the diffusion region, this motion speed is translated 
to a generic reconnection rate in terms of the Alfv\'en Mach number $M_a = v_{in}/v_{a} \sim 0.001  - 0.1$ \citep{Isobe2005}. 

The apparent motion in the third dimension breaks the translational symmetry of the 2d model. 
It is entirely plausible that the 3d magnetic topology dictates the allowed flare connectivity 
and therefore the geometry of flare ribbons and loops \citep{Aulanier2006}. The question is one of dynamics: 
what mechanism determines the speed of such spreading of flare brightening or,
alternatively, what physical properties characterize magnetic reconnection?
A few different scenarios have been proposed to discuss the nature of this motion. They
can be grouped into two categories.


Linear MHD waves have been invoked to explain the propagation of perturbations. 
\citet{Vorpahl1976} proposed that magneto-acoustic waves are responsible for the sequential brightening of X-ray coronal 
loops along the axis of the arcade. If the axis of the arcade lies in the $y$ direction, 
the velocity component in this direction is given by $$2v_y^2 = v_s^2 + v_a^2 \pm \sqrt{(v_s^2+v_a^2)^2 -4v_s^2v_a^2\cos^2\phi}, $$
where $v_s$ and $v_a$ are the local sound speed and Alfv\'en speed, respectively, 
and $\phi$ is the angle between the axis of the arcade and the magnetic field vector. We call
the $y$-component of the magnetic field the guide field $B_g$, so that $\cos\phi = \hat{B}\cdot\hat{y}
= B_g/B$, where $B$ is the total magnetic field strength. The fast mode along the $y$ direction spreads at 
a speed greater than Alfv\'en speed $v_a$ calculated with the total strength of the magnetic field
$B$. \citet{Vorpahl1976} assumed that the magnetic guide field $B_g = 0$, and estimated the speed of the 
fast magnetosonic waves $v_y = \sqrt{v_a^2+v_s^2}$ to be a few times the observed speed of arcade 
spreading at 180 -- 280 km s$^{-1}$. 

With the presence of guide field $B_g \neq 0$, an Alfv\'en wave can also be supported traveling along
the $y$ axis \citep{Katz2010}, at the Alfv\'en speed calculated using the guide field $$v_y = v_{a,g} = \frac{B_g}{\sqrt{\mu_0 \rho}} 
= 700 \frac{B_{g, 1}}{\sqrt{n_9}}~ {\rm km~s^{-1}}, $$ where $\rho$ is the plasma mass density, $B_{g, 1}$ 
is the guide field in units of 10$^1$ G, and $n_9$ is the electron number density in units of 10$^9$ cm$^{-3}$. Since
the Alfv\'en speed is determined by only a component of the coronal magnetic field, $v_y$ could be smaller than the speed of magnetosonic waves by a factor of a few. 

In these MHD wave scenarios, $v_y$ as estimated above is of order 10$^3$ km s$^{-1}$ given the general coronal 
properties. It should also be noted that MHD waves spread bi-directionally and symmetrically along the axis of the arcade. 

Another macroscopic scenario based on a 3-dimensional framework suggests that 3d reconnection 
along Quasi-Separatrix Layers \citep[QSLs, ][]{Priest1995} results in spreading of the ribbon \citep{Aulanier2006}. 
3d MHD numerical simulations have been conducted to qualitatively demonstrate the observed motion of flare kernels or loops
\citep{Masson2009, Aulanier2012, Janvier2013, Dudik2014, Savcheva2015}. Some recent numerical simulations show
the evolving QSLs ``resulting from the expansion of a torus-unstable flux rope" \citep{Aulanier2012, Dudik2014}, which
might qualitatively explain the relationship between filament/CME eruption and propagation of flare loops observed
by \citet{Tripathi2006, Liu2010}. It is not clear what controls the speed of the apparent motion in the topology models.

A different mechanism for ribbon spreading has been proposed due to microscopic current dynamics,
motivated by observations in magnetospheric substorms, for which the guide field is typically small. It was proposed that 
drifting of current carrying particles leads to spreading of reconnection
along the current sheet, in the direction of the species carrying the current \citep{Huba02,Huba03,Shay03,Karimabadi04,Lapenta2006}. 
The spreading speed is $$v_y = \frac{B_{rec}}{\mu_0 ne\delta},$$ where $B_{rec}$ is the reconnecting component of the magnetic field,
$n$ is the electron number density in the current sheet, $e$ is the charge, and $\delta$ is the thickness of current sheet. This 
can be further expressed as $$v_y = 5.0 \frac{B_{rec, 1}}{n_9\delta_5}~{\rm km~s^{-1}}, $$ where $B_{rec,1}$
is in units of 10$^1$ G, $n_9$ in units of 10$^9$ cm$^{-3}$, and $\delta_5$ in units of 10$^5$ cm. 
The thickness $\delta$ is related to the microphysics allowing reconnection.  While it is unlikely that classical Sweet-Parker reconnection occurs (due to collisonless effects and secondary islands), we employ the Sweet-Parker model to get an estimate for the spreading speed.  In the Sweet-Parker model, $\delta \sim LS^{-1/2}$, where $L \simeq 10^{9}\,{\rm cm}$ is the global scale of the current sheet.  
A typical Lundquist number, $S \sim 10^{12}$, gives $\delta_5 \sim 0.01$. This yields an estimated $v_y$ of at least 500 km~s$^{-1}$, which can be comparable with the propagation speed of MHD waves. 
Nevertheless, in this mechanism, $v_y$ is uni-directional; if electrons are the current carriers, $v_y$ is anti-parallel 
to the direction of the current $\vec{J}$ in the current sheet.

Very recently, \citet{Shepherd2012} conducted a theoretical analysis as well as numerical simulations 
to examine propagation of reconnection along a current sheet as a function of varying guide field. They argued that the spreading is controlled by whichever of the two mechanisms is faster, so that
reconnection spreads bi-directionally at the Alfv\'en speed in a current sheet with a strong guide field $B_g$,
asymmetrically with a weak guide field, or
uni-directionally by current carrying particles with a zero guide field. 

As a result of the experimental, numerical, and theoretical work, it therefore appears that observations of elongation motion of flare ribbons may be used to infer properties
of the coronal current sheet where reconnection takes place, such as the magnetic guide field or even the thickness of the current sheet.
These properties are not directly measurable, but they are critical to our understanding of reconnection dynamics and energetics in solar flares. To be able 
to make an association between observations and models, the configuration of the flare needs to be relatively simple so that we 
can define the translational direction. For such a purpose, we analyze two-ribbon flares taking place along a roughly straight magnetic 
polarity inversion line in active regions dominated by a bipolar configuration. We measure the apparent motion 
of ribbon brightening along the polarity inversion line, the presumed translational direction, as well as the inclination 
of flare loops with respect to the polarity inversion line to estimate the magnetic guide field relative to the total 
field of the flare loop. These measurements are then compared to examine whether the speed and direction of elongation are related to the  
shear to allow us to infer the current sheet properties. In the following text, the method of analysis will be described in Section 2 and applied
to three two-ribbon flares in Section 3. To enlarge the sample, we also include, in Section 4, results of other events analyzed previously.
In Section 5, properties of ribbon motion are compared with magnetic properties of flare regions in search for understanding about governing mechanisms of ribbon elongation. Conclusions and discussions are given in the last section.

\section{OBSERVATIONS AND ANALYSIS}

This study will present analysis of three two-ribbon flares to illustrate apparent motion 
patterns of flare ribbon brightening. 
These events are selected because they demonstrate ordered ribbon motion along rather collimated stretch of ribbons, 
allowing a 2.5d approximation and identification of the direction of the magnetic polarity inversion line as the translational
direction, along which the ribbon spreading speed can be measured. In addition to comprehensive analysis of these three events,
we will also review a few more flares previously analyzed and published, and summarize their properties together with the three cases.

The general information about these flares is given in Table~\ref{flareinfo} (Event 1, 2, and 3, respectively). 
All three flares analyzed in this study occur near the disk center in nearly bipolar magnetic configurations. The first event, C-class 
flare SOL2011-09-13, takes place in a growing active region consisting of a sunspot and a plage. The active region 
does not have a filament. For many hours before the flare, a soft X-ray sigmoid consisting of an arcade of sheared loops is 
visible in XRT onboard Hinode (McKenzie, private communication) as well as in the 335~\AA\ passband of the {\it Atmosphere Imager
Assembly} \citep[{\it AIA}, ][]{Lemen2012}. Disruption of these loops is observed at the onset of the 
flare, and a new arcade of loops formed afterwards. This flare was also reported by \citet{Hu2014, Qiu2016}. The 
second event, a C5.7 flare SOL2011-12-26, occurs in an active region without major sunspots. 
A remnant piece of a filament is visible in the active region; however, the filament 
is not observed to erupt, though a CME is observed by the {\it Solar TErrestrial RElations Observatory}({\it STEREO})
and analyzed by \citet{Cheng2016}. The third event, an
M8.0 flare SOL2005-05-13, occurs in an active region consisting of a sunspot and a plage. A filament 
sits along the magnetic polarity inversion line, and is partially erupted during the flare 
\citep{Qiu2005, Yurchyshyn2006, Liu2007, Tripathi2009, Kazachenko2009, Liu2013}. The first two flares were observed by AIA, and 
the last by {\it TRACE}.

These flares exhibit two ribbons well-observed in the UV 1600 \AA\ passband on both sides of the PIL, 
with the shape of the ribbons outlining the PIL. Seen in Figures~\ref{0913_1}, ~\ref{1226_1}, and ~\ref{0513_1}, in each 
flare, the ribbon brightening starts locally and then spreads along the PIL to form the full length of the ribbon 
 -- we call this motion the {\em elongation} motion. In some of these events, because of the asymmetry 
of magnetic fields of opposite polarities, the elongation motion is 
not symmetric on the two ribbons, with brightening of one ribbon in the weaker magnetic field moving faster than the other in
stronger field so as to balance the positive and negative reconnection flux. The more well-known apparent motion 
of the flare ribbons, the perpendicular expansion away from the PIL, also often referred as the {\em separation} 
motion, usually dominates after the elongation. In the framework of the 
two-dimensional standard model, this separation motion has been related to the inflow speed at the boundary 
of the diffusion region in the corona, and therefore has been used to measure mean reconnection rates in terms of the 
macroscopic electric field in the reconnection current sheet \citep{Poletto1986, Qiu2002, Isobe2002}, although it
remains unclear what exactly determines the electric field, or the rate of fast reconnection. 
The physical meaning of the first motion, the elongation, is much less understood, and this motion is the focus of this paper.

As in \citet{Qiu2009, Qiu2010, Cheng2012}, we decompose the apparent motion of the front of the ribbon brightening into 
components along and perpendicular to the PIL. For each active region, the shape of the PIL relevant to the flare
is outlined semi-automatically based on the pre-flare line-of-sight magnetogram and is then fitted by a polynomial, as 
shown with the orange symbols and curve, respectively, in Figures~\ref{0913_1}, ~\ref{1226_1}, and ~\ref{0513_1}. At a given moment of the flare, the trajectory of the 
outer-most pixels of the ribbon is projected onto the PIL to define the positions of the two ends of the newly brightened ribbon 
front along the PIL, denoted by $l_w$ and $l_e$ for the two ends in the west and east, respectively. The length $l_{||}=|l_w-l_e|$ 
along the PIL between the two ends is also measured. The area encompassed by the newly brightened ribbon pixels and the PIL 
is then divided by this length $l_{||}$ to yield the measurement of the mean perpendicular distance of the ribbon to the PIL, denoted by $\langle d_{\perp}\rangle$. 
The measurements have been made using varying thresholds of UV counts (as how many times the median or quiescent 
UV counts) to define ribbon fronts, and the standard deviation of these measurements is given as the uncertainty (error bars in the relevant plots).
With these measurements, the apparent ribbon motion speed is found, $v_w = \Delta l_w/\Delta t$, $v_e = \Delta l_e/\Delta t$,
and $\langle v_{\perp}\rangle = \Delta \langle d_{\perp}\rangle/\Delta t$. In a strictly two-dimensional version of the standard model, 
$v_w = v_e = 0$, and reconnection in the corona is characterized by a macroscopic electric field along the 
current sheet, assumed to be in the direction of the PIL, $\vec{E} = - \vec{v}_{in} \times \vec{B}_{in} \approx 
\vec{v}_{\perp}\times\vec{B}_n$. 
In practice, it has been assumed that $\vec{B}_n$ does not change significantly from
the photosphere to the chromosphere, nor during the flare, so that $\vec{B}_n$ is usually taken from
conventional photospheric magnetograms such as those provided by MDI and HMI. For active regions near disk center, it
is also often approximated that $\vec{B}_n$ is the same as the longitudinal magnetic field.

To examine the possible relation of the ribbon motion to properties of the coronal current sheet, we also note the direction 
of the elongation motion with respect to the direction of the macroscopic electric current in the coronal current sheet.
When a potential field anchored at photospheric bipoles is stretched upward, an electric current may be produced and
its direction is determined by Ampere's law $\vec{J} \propto \vec{\nabla} \times\vec{B}$. Whereas the exact magnetic field 
configuration or field strength is not known in the corona, the directionality of this field is reasonably deduced from 
photospheric magnetograms. In this way, the direction of the current density $\vec{J}$ is marked
in the figures for the studied active regions, and the ribbon elongation motion is referred as ``parallel", ``anti-parallel", or ``bi-directional" 
with respect to this direction\footnote{Some previous studies refer ``parallel" or ``anti-parallel" motion as conjugate foot-points
moving in the same or opposite directions, different from the definition used in this study.}. 

We also report the inclination angle of post-reconnection loops with respect to the PIL.
In a 2.5d approximation, the tangent of this angle would indicate the ratio of the magnetic outflow field $B_o$
and the magnetic guide field $B_g$, the component along the current which does not participate in reconnection.
This inclination angle is measured in two ways. The first way, used in in many previous studies, 
measures the angle between the line connecting the two ribbon fronts and the PIL determined from the photospheric
magnetogram. When flare loops are observed, we also outline these loops and compute the angle between the loop and the PIL 
(determined from a potential field extrapolation) where they cross. To compare flare loops with a potential field, 
we extrapolate a potential field and trace field lines from the same photospheric points as the potential-field
loops, and find the angle where the projected field lines cross the PIL.

Finally, it should be noted that the emphasis of this study is to illustrate the global tempo-spatial pattern of ribbon brightening.
A flare ribbon usually consists of a series of bright kernels but is {\em not} a continuous smooth patch of emission, as
is also the case for the events in this study. According to current belief, these kernels are the feet of
loops newly formed by reconnection. Energy released by an individual reconnection event propagates downward along the loop
and heats the chromosphere at the foot-point kernels. It is also assumed that the foot-points of a loop 
do not move (i.e., line-tied assumption) over the flare timescale, hence apparent motion of flare ribbon brightening refers 
to successive energy release and atmospheric heating at adjacent locations, 
and these brightenings map successive reconnection events in the corona down to the chromosphere.

\section{APPARENT MOTION OF FLARE RIBBONS}

\subsection{Persistent Anti-parallel Elongation (2011-09-13)}
\begin{figure}
\plotone{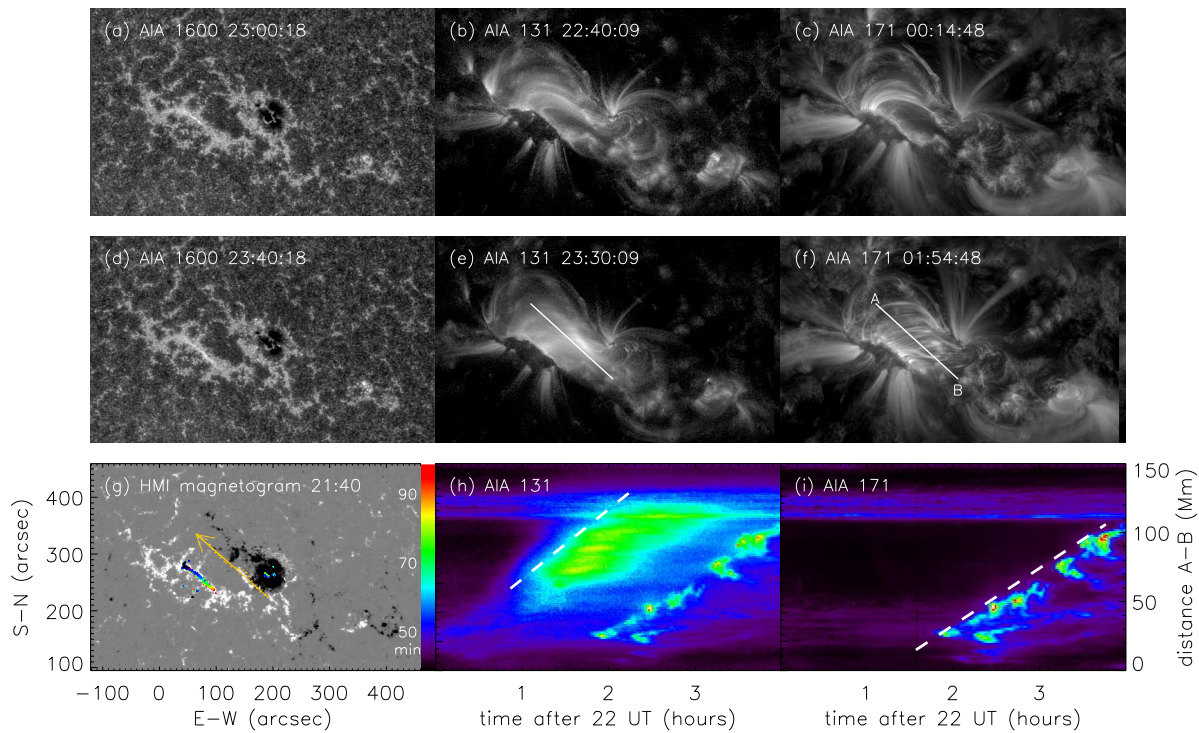}
\caption{(a-f): snapshots of the flare SOL2011-09-13T22 observed at three passbands of {\it AIA}
during its evolution. (g) Longitudinal magnetogram (greyscale) by HMI superimposed with the positions
of newly brightened ribbons (color). Time in minutes from 22:00 UT is indicated by the color code.
The orange curve outlines the PIL of the photospheric longitudinal magnetogram, and the arrow indicates the direction
of the macroscopic electric current in the corona.
(h) Time-distance stack plot of loop top emission in the EUV 131 passband along the axis of the flare arcade (indicated by the
solid white line in panels (e) and (f)). The dashed guide line outlines the front of spreading loops
at an average speed of 13~km s$^{-1}$. (i) Time-distance stack plot of loop top emission in the EUV 171 passband 
along the axis of the flare arcade. The dashed guide line outlines the front of spreading loops
at an average speed of 10~km s$^{-1}$. } \label{0913_1}
\end{figure}

\begin{figure}
\plotone{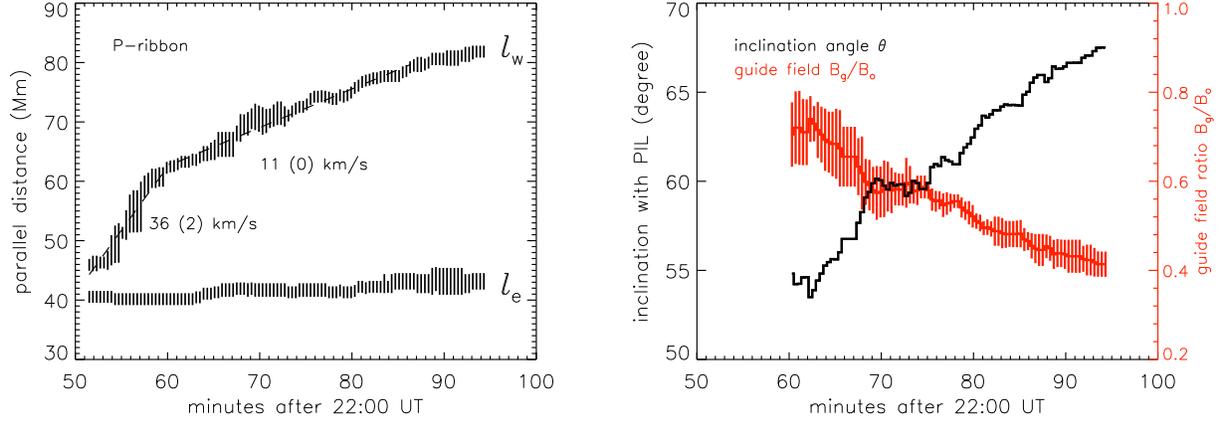}
\caption{Left: position of the newly brightened front of the UV ribbon in the positive magnetic field along the PIL. Right: the inclination angle
$\theta$ of the line connecting conjugate flare foot-points with respect to the PIL (black), and the presumed ratio of the guide field
to the outflow field $B_g/B_o = {\rm cot}\theta$.}\label{0913_2}
\end{figure}

The C-class flare SOL2011-09-13T22:00 exhibits persistent  elongation motion of the positive ribbon front, with little
perpendicular expansion following the elongation. The left panels in Figure~\ref{0913_1} show the 
flare morphology in the UV 1600 \AA\ band observed by {\it AIA}, and mapping of flare ribbons
on a line-of-sight magnetogram obtained by HMI. Elongation of the ribbon in the plage of the positive 
magnetic field proceeds for nearly an hour. The distance of the ribbon brightening front along 
the PIL in both directions is measured and shown in Figure~\ref{0913_2}. The brightening spreads only in 
one direction (i.e., only $l_w$ grows) at a mean speed of 36 km~s$^{-1}$ in the first 5 minutes, and then at 11
km~s$^{-1}$ for the following hour. We note that the extension and motion of the ribbon in the sunspot of negative 
magnetic field is insignificant because of asymmetric magnetic field configuration at the two polarities, so the motion is not measured in the negative ribbon. 

The ribbon brightening is then followed by sequentially formed flare loops first showing up in the northeast, 
and then spreading down along the PIL. Figure~\ref{0913_1} (b) and (e) show 
the arcade evolution in the {\it AIA} 131 passband characterized by plasma emissions at 10~MK, and Figure~\ref{0913_1} (c) and (f)
show evolution of the flare loop arcade in the 171 passband of plasma cooled 
to 1~MK in about 90 minutes \citep{Qiu2016}. The speed of the apparent spread of the loops can be estimated 
from the stack plots along the axis of the arcade. The slit AB crossing the loop tops is shown in panels (e) and (f), and the stack plots along AB for the two passbands, 
of the hot 131 emission and cool 171 emission, are shown in panels (h) and (i), respectively. The two dashed guide lines
suggest the loop spreading speed at 14 and 10 km s$^{-1}$, respectively, following emission fronts at the two passbands. 
The measurements have also been made with other passbands and along slits crossing different parts of the loop arcade from
the legs to the top, by either tracking the front or the maximum brightness of the time-distance stack-plot. In four passbands
131 (10~MK), 94 (6~MK), 335 (3~MK), and 171 (1~MK), measurements tracking the front yield 
a mean speed of $13.9\pm 0.7, 11.2\pm1.6, 9.8\pm1.8, 9.6\pm0.5$ km s$^{-1}$, respectively. 
Measurements tracking the peak brightness give the mean speed of $7.0\pm0.5, 6.2\pm0.9, 6.8\pm2.9, 9.6\pm0.3$ km s$^{-1}$, which are systematically
smaller than the first measurements except for the 171 \AA\ band. In short, the pattern of loop spreading is consistent with ribbon spreading,
although the measured speed of the loops is slightly lower than ribbon spreading.
 
The bipolar magnetic field configuration in this event allows us to determine the direction of the electric current of
stretched bipolar field at the possible location of reconnection above the PIL, which is indicated by the arrow in Figure~\ref{0913_1}(g). In this event,
the ribbon/loop spreading is anti-parallel with respect to the current $\vec{J}$. Interestingly, the event exhibits
little perpendicular expansion, and we therefore do not measure $v_{\perp}$ here.

With the 2.5d approximation as illustrated by the arcade configuration, we may estimate the inclination angle $\theta$
of flare loops with respect to the PIL. The complementary angle of $\theta$ has been referred as 
the shear angle in previous studies. Therefore, a smaller $\theta$ in this study refers to a stronger shear as defined
in previous studies. The right panel of Figure~\ref{0913_2} shows evolution of the 
(foot-point) inclination $\theta_f$ defined as the angle between the line connecting conjugate foot-points 
(the brightening fronts in the two ribbons) and the PIL of the photospheric magnetogram. This angle 
gradually increases from 55$^{\rm o}$ to 70$^{\rm o}$ as ribbons and loops spread along the PIL\footnote{Measurement of $\theta_f$ before 23:00~UT is not presented since it 
carries large uncertainties due to weak emission in the negative field of the sunspot.}, 
which is consistent with many previous reports of decreasing shear during the flare evolution. 

Since both pre- and flare loops are observed in the EUV wavelengths, we also demonstrate 
the inclination angle of the observed loop top with the PIL of the coronal magnetic field,
which is approximated by a potential field extrapolation using the photospheric magnetogram as the bottom boundary.
This inclination angle is denoted by $\theta_b$ and $\theta_a$, referring to loops before 
and after the flare, respectively. 

\begin{figure}
\plotone{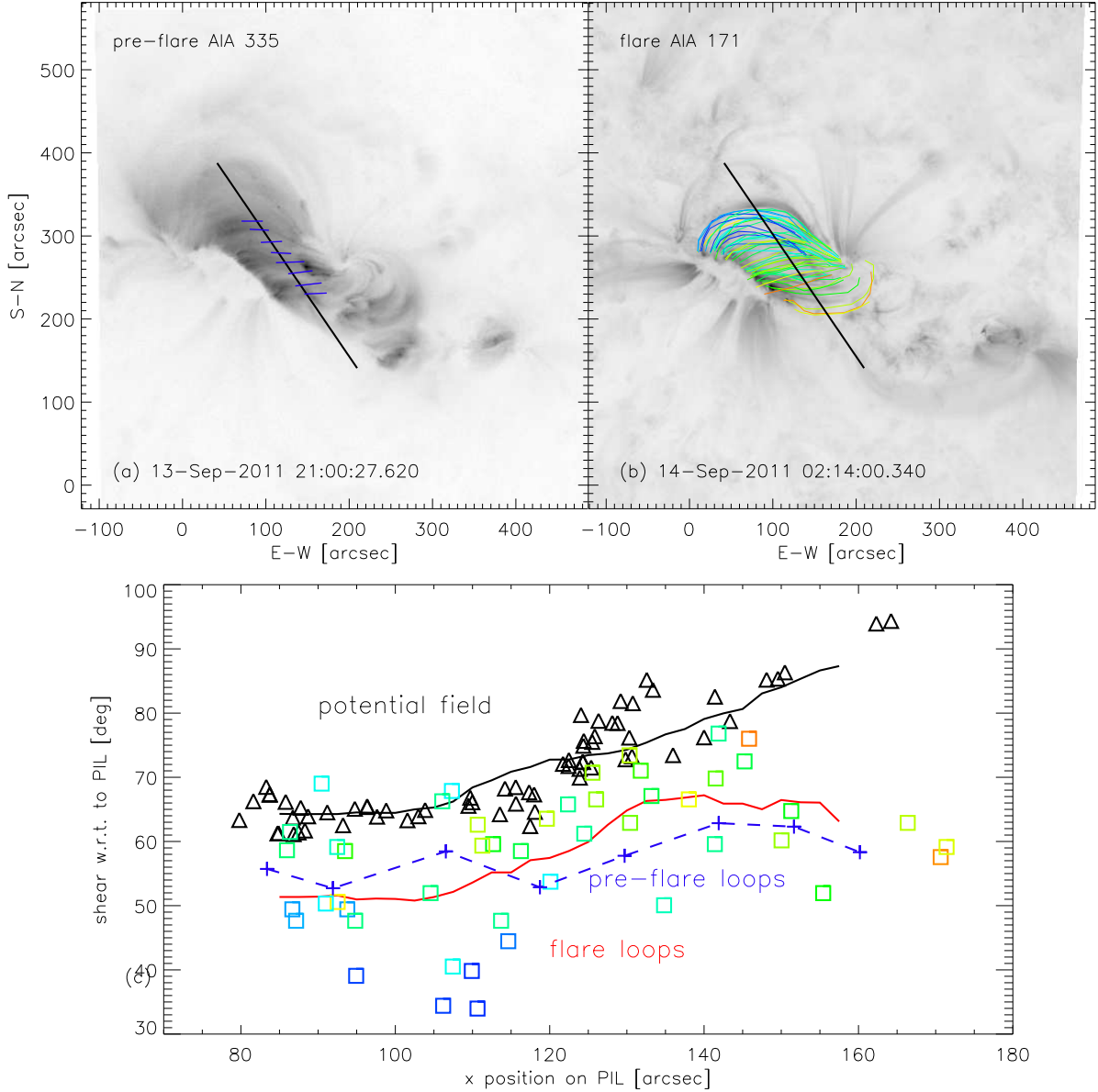}
\caption{Top left: pre-flare loop arcade at 21:00 UT observed in the {\it AIA} 335 passband, with the loop tops outlined in blue bars, and 
the black solid line indicating the magnetic PIL at the estimated height of the loop top. Top right: flare loop arcade at 02:14 UT observed 
in {\it AIA} 171 passband, with sequentially formed flare loops outlined in color curves (cold color loops form earlier
than warm color loops), and solid line indicating 
the magnetic PIL at the estimated height of the loop top. 
Bottom: the inclination angle of the pre-flare ($\theta_b$, blue dashed line) and flare loops ($\theta_a$, color symbol) along the PIL.
The red curve shows the average loop inclination along the PIL. Black symbols show the inclination angle $\theta_p$ of potential field lines anchored at
the ribbon pixels, and the black curve shows the average potential field inclination along the PIL.
}\label{0913_3}
\end{figure}

\begin{figure}
\plotone{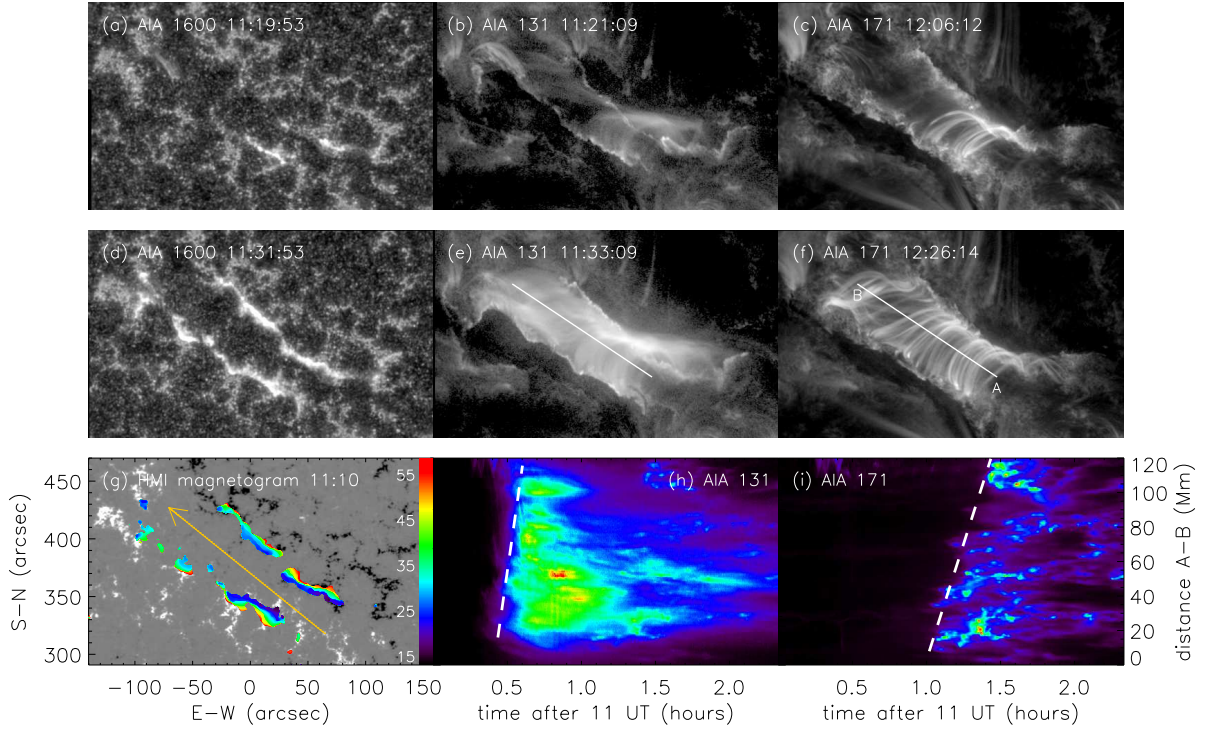}
\caption{Same as Figure~\ref{0913_1}, for the flare SOL2011-12-26T11:23. The guideline in panel (h) suggests spreading of loops
in the 131 passband at the mean speed of 110 km s$^{-1}$, and that in panel (i) indicates spreading of loops in the 171 passband at the mean speed of
50km s$^{-1}$.} \label{1226_1}
\end{figure}

Pre-flare loops are visible in the 335 \AA\ passband, suggesting that
the typical temperature of the pre-flare arcade is about 3~MK. A set of loops is selected manually along the axis of 
the arcade observed in a 335 \AA\ image taken at 21:00~UT, about 1.5 hours before the flare onset. This is 
shown in Figure~\ref{0913_3}a. The estimated inclination angle $\theta_b$ of these loops along the coronal
PIL, illustrated by the black solid line, is plotted in the blue dashed line in Figure~\ref{0913_3}c.
$\theta_b$ is estimated to vary between 50 and 60 degrees along the PIL in the corona.

Flare loops form sequentially and are visible in multiple passbands (see Figure~\ref{0913_1}).
They are well observed in the 171 passband with the best contrast. About 100 of these loops in the 171 images, 
formed at different times at different locations along the PIL, are outlined manually, and a subset of them 
is illustrated in Figure~\ref{0913_3}b. Color coding indicates the time when a 
loop is first visible in this bandpass, with violet and blue loops formed earlier than yellow and orange loops. 
Note that these loops start to appear more than an hour after the ribbon brightening due to elongated plasma heating
and cooling in the corona \citep{Qiu2016}. We estimate the angles made by the top of these flare loops with the 
PIL of the coronal magnetic field. Color symbols in Figure~\ref{0913_3}c
show the inclination angle of these loops formed at different times along the PIL. It is shown that earlier formed loops (cold colors)
carry a larger shear (smaller $\theta_a$) of down to 40$^{\rm o}$. Later formed loops (warm colors) have decreased
shear with $\theta_a$ up to 75$^{\rm o}$. The solid red curve shows the average inclination angle
$\theta_a$ along the PIL, which grows westward as the flare ribbon and loops spread westward. 
Such an evolution trend and the range of inclination angles are consistent with the results from the foot-point measurements.
The inclination angle of most flare loops is also larger than that of pre-flare loops.

In a 2.5d configuration, an inclination angle different from 90$^{\rm o}$ indicates the presence of a guide field $B_g \neq 0$, but
does not necessarily indicate a non-potential magnetic loop. To compare the estimated
inclination angle of pre-flare and flare loops with the potential field itself, we also
calculate the angles of potential field lines anchored at the flare ribbons at the points their projections cross the PIL. These angles ($\theta_p$) are plotted in black symbols in Figure~\ref{0913_3}c. The angle $\theta_p$
also grows during the westward propagation, but the potential field loops generally make a larger angle than flare loops.
The comparison therefore suggests that, on average, flare loops are more sheared than the potential field
and less sheared than the pre-flare arcade. 

The flare loop inclination may be translated into the ratio of the guide field to the outflow field 
by the relation $B_g/B_o = {\rm cot}\theta$, assuming that $\theta$ is the same as $\phi$, the angle 
of magnetic field line with the translational direction. In this event, at the start of the flare, the inclination angle
is down to 40$^{\rm o}$, when the elongation speed is 36 km s$^{-1}$. Afterwards,
the average flare inclination varies from 50$^{\rm o}$ to 70$^{\rm o}$, and hence $B_g/B_{o} \approx 0.8 - 0.4$, 
as the flare evolves with a steady ribbon elongation at 11 km s$^{-1}$
for about an hour.

In summary, we find this event is a special example of slow and persistent spreading of
reconnection along the PIL with a relatively weak shear. Note that this flare does not exhibit significant 
expansion of the ribbon perpendicular to the PIL, suggesting that the reconnection site might not rise in the corona
as typically assumed in the standard flare model. It is notable that a potential-field source-surface (PFSS) model 
with a source-surface at $2.5\,R_{\odot}$, reveals this AR to have several unusual properties. It lies squarely 
underneath a closed separatrix dome whose null point lies practically at the source surface: $r=2.3\, R_{\odot}$.  
It is possible that an eruption opened this dome, which then reformed through reconnection unusually high in the 
corona, which proceeded unusually slowly, with little notable outward expansion.

This example reveals several other properties  
not seen in previous studies. Most remarkably, the measured shear and its evolution from the foot-point brightenings
with respect to the photospheric PIL is largely consistent with the measurement of flare
loops with respect to the coronal PIL. It is also found that the motion pattern derived from the UV 
ribbon brightening is consistent with that derived from flare loops. Despite the asymmetry 
in the motion of conjugate foot-points, flare arcade exhibits a rather homogeneous motion, 
suggesting that the coronal magnetic field is indeed more homogeneous than the photospheric distribution.
On the other hand, the measured speed of flare loops is systematically lower than the 
speed of the foot-point motion, suggesting that the timescale of coronal loop spreading may be 
convolved with hydrodynamic timescales due to plasma heating and cooling in the corona. 

\subsection{Fast Parallel Elongation (2011-12-26)}
In the next example, we observe elongation of flare ribbons parallel to the direction inferred for the
electric current along the PIL. The two-ribbon flare SOL2011-12-26T11:23 exhibits two nearly symmetric flare ribbons, both of which start to brighten
from the southwest end of the AR and spread along the PIL toward the northeast. The left panels of 
Figure~\ref{1226_1} show the evolution of the two ribbons; in both ribbons, the brightening spreads 
along the PIL in one direction (i.e., only $l_e$ grows) for about 10 minutes at the mean speed of 81 
and 98 km s$^{-1}$ for the positive and negative ribbons, respectively. These speeds are an order of 
magnitude greater than in the first event. 

Again, for this flare, the flare arcade is observed in multiple wavelengths, allowing us to
measure the speed of the spreading loops. The panels in the middle column show evolution of
the flare loops in the 131 passband and the stack plot along a slit crossing the loop tops
along the axis of the arcade, and the right column shows observations and measurements in the 171 passband.
For this event, the measured speed of spreading hot (131) loop fronts is very similar to the speed of the ribbon
spreading, of 110 km s$^{-1}$; however, at the low temperature (1~MK) passband, the apparent motion of the cool loop
front is a lot slower, at only 50 km s$^{-1}$, suggesting that timescales of hydrodynamic evolution (e.g. cooling)
of loop plasmas have significantly affected the apparent speed of spread.

The active region hosting the two-ribbon flare is primarily bipolar with nearly symmetric positive and negative fields, 
both located in plages. The direction of the electric current is derived and indicated in the bottom left panel in Figure 4.
The ribbon elongation is parallel to $\vec{J}$, which is different from the first event. 

Also different from the first event, in this flare, the ribbon separation 
motion is observed, and the mean perpendicular speed in the positive ribbon is about 10 km s$^{-1}$ in 
the first 10 minutes. Afterwards, as the two ribbons have fully developed and elongation motion has stopped, 
the separation motion continues at a mean speed of about 2 km s$^{-1}$ in both ribbons for another 30 minutes.

\begin{figure}
\plotone{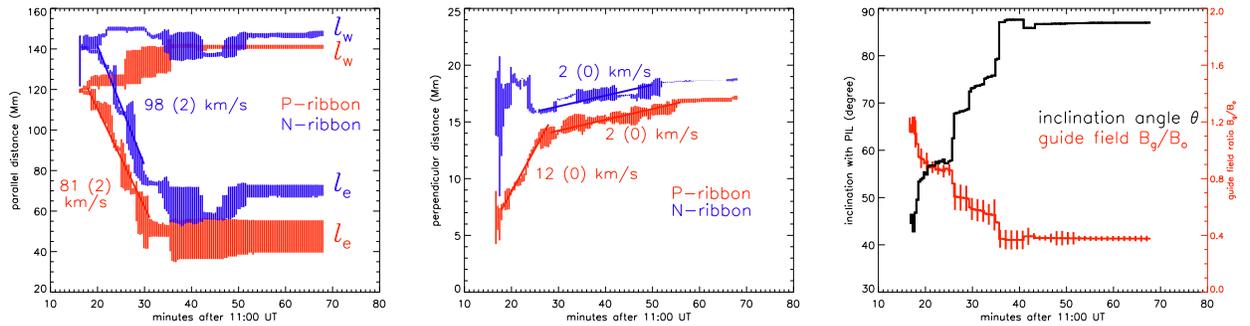}
\caption{Left: positions of the newly brightened fronts of the two UV ribbons along the PIL. Middle: the mean distance of the UV ribbon
fronts from the PIL. Right: the inclination angle $\theta$ of the line connecting conjugate flare foot-points with 
respect to the PIL (black), and the presumed ratio of the guide field to the outflow field $B_g/B_o = {\rm cot}\theta_f$.
} \label{1226_2}
\end{figure}

In a similar manner, we estimate the inclination angle from newly brightened conjugate foot-points.
Figure~\ref{1226_2} shows that this angle $\theta_f$ changes from 45$^{\rm o}$ to nearly 90$^{\rm o}$ within 20 minutes, 
reflecting a more significant shear variation compared with the first event. From this angle, the estimated 
$B_g/B_o$ varies from 1 at the start of the flare to 0.

\begin{figure}
\plotone{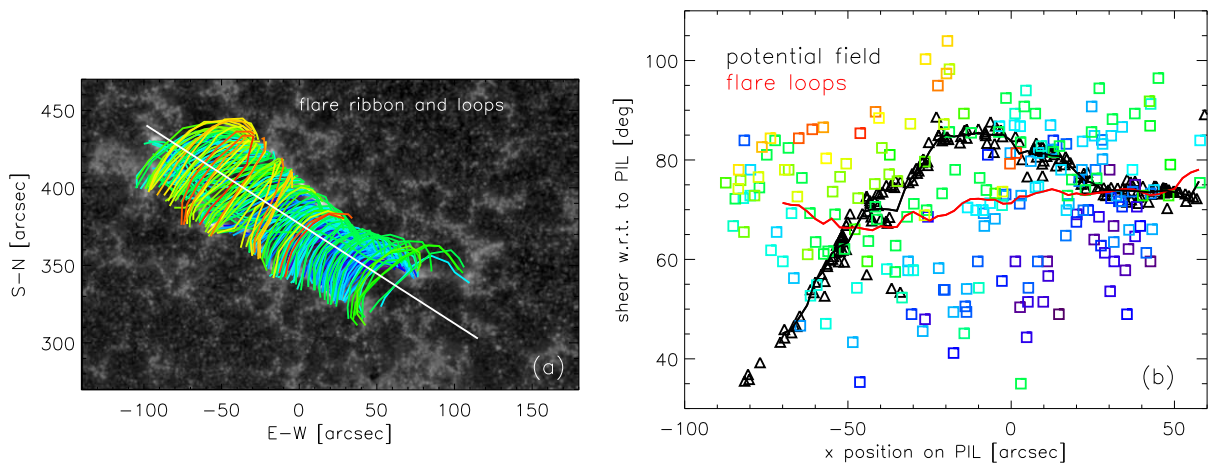}
\caption{Left: flare loops in the EUV 171 bandpass formed sequentially along the PIL, superimposed on the UV 1600 image showing
flare ribbons at the feet of the loop arcade. The white solid line indicates the PIL at the estimated height of the loop top.
Right: inclination angle of the top of the loops shown in the left with the PIL at the estimated height of the loop top, 
as in Figure~\ref{0913_3}c.}  \label{1226_3}
\end{figure}

\begin{figure}
\plotone{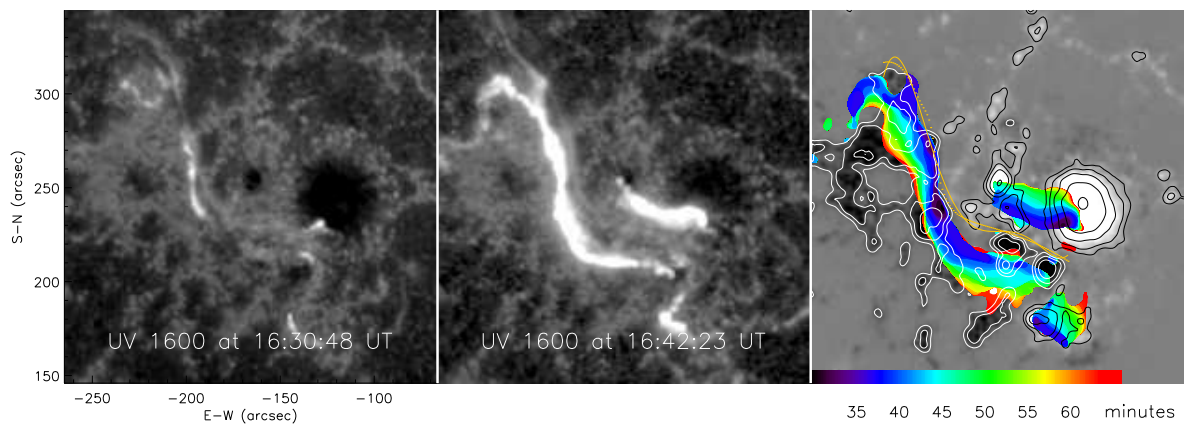}
\caption{Evolution of the two ribbons of the SOL2005-05-13T16:35 flare observed in the UV 1600 bandpass of {\it TRACE}. In the right panel, the positions
of newly brightened ribbon fronts are superimposed on a longitudinal magnetogram by MDI, with the time lapse given by the color code.} 
\label{0513_1}
\end{figure}

Shear angles can be also estimated with flare loops.
More than two hundred flare loops are manually outlined in the 171 images and shown in Figure~\ref{1226_3}a. 
The right panel shows the measurement of inclination along the PIL of the coronal potential field, 
with cold colors indicating loops formed earlier than warm-color loops. In this event, because loops form
quickly along the PIL in the initial phase and then ``grow" upward in a nearly 2d manner, the measurement
of inclination along the PIL does not seem to show a clear pattern due to mixture of earlier and later formed loops
at the same PIL position, but there is clear indication that earlier formed (cold-color) loops are, on average,
more sheared, $\theta_a \approx 45 - 65^{\rm o}$, than later formed (warm color) loops with $\theta_a \approx 60 - 90^{\rm o}$.
Again, the range and evolution of the inclination of flare loops are remarkably consistent with the foot-point measurements. 

The inclination angle of potential field loops anchored at the ribbons is also estimated. In the west where
ribbon brightening starts, flare loops are more sheared than potential loops; in the east, however, the small inclination
angle of the potential field may be caused by the changed height of loops, where the orientation of the PIL may have changed.

This flare is associated with a partial halo-CME well observed by {\it STEREO} \citep{Cheng2016}. The CME is first detected at
11:06 UT, and appears to rise at 11:10 UT when brightening of UV ribbons starts. An analysis by \citet{Cheng2016} shows that
CME height evolves with reconnection measured in ribbons. The {\it AIA} disk observations show a faint remnant of a low-lying filament
which did not erupt. Neither did the {\it STEREO} 304 images show filament eruption prior to the CME.
It is observed that some activation of the filament remnant, including brightening and flows along the axis, is present at the northeast end
of the filament, yet flare ribbon brightening starts at the southwestern end. 
So it is not clear whether and how the global dynamics associated with the CME eruption would govern the ribbon elongation.

\subsection{Fast bi-directional elongation (2005-05-13)}
\begin{figure}
\epsscale{1.0}
\plotone{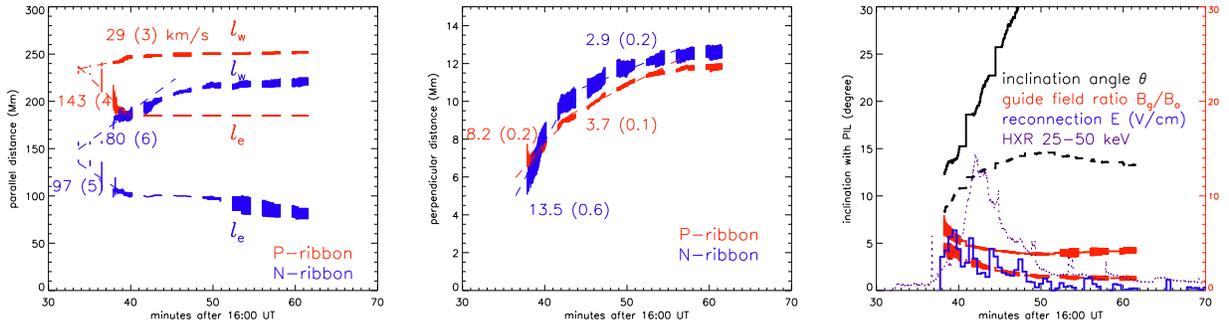}
\caption{Left: elongation of the two ribbons in both the west and east directions along the PIL. Middle: mean distance of the two ribbons from the PIL. Right: inclination
angle of the flare (see text), the presumed $B_g/B_o$, the mean electric field given by $\langle E \rangle = v_{\perp} B_n$, and the HXR 25-50 keV lightcurve. } \label{0513_2}
\end{figure}

The third event is a well-studied M8.3 two-ribbon flare SOL2005-05-13,
which has been analyzed by \citet{Yurchyshyn2006, Qiu2005, Liu2007}, as well as modeled
by \citet{Kazachenko2009, Liu2013}. The flare was observed 
by {\it TRACE} mostly in the 1600 \AA\ bandpass with a very high cadence of 3~s when the flare mode was switched on. Figure~\ref{0513_1} shows
evolution of the two ribbons. Measurements of ribbon motion are plotted in Figure~\ref{0513_2}. For 
this flare, ribbon brightening starts at the middle in both ribbons, and spreads in both directions 
along the curved PIL. Spreading of the ribbon in the weak and negative magnetic field proceeds nearly symmetrically
in both directions with an average speed of $v_w \approx v_e \approx 100~ {\rm km~s}^{-1}$. 
In the strong positive magnetic fields inside a sunspot, the ribbon spreads asymmetrically; it moves 
more quickly towards the penumbra with $v_e \approx 80$ km ~s$^{-1}$, and in the opposite direction into the 
umbra, the ribbon spreading is much slower $v_w \approx 20$ km ~s$^{-1}$.
We recognize that there are large uncertainties in the measurements of the speed because
of the very low cadence (80 -- 170 s) of {\it TRACE} observations at the start of the flare, when we only
have 2 -- 3 measurements as ribbon spans over a long distance.
Because of the low cadence, it is also seen that when the ribbon starts to brighten, an extended 
section, rather than a compact kernel, of the ribbon is brightened. Nevertheless, it can be observed that ribbon spreading is 
bi-directional from both ends of the section at a relatively high speed of order 100 km~s$^{-1}$.

In this event, perpendicular expansion of the ribbons is also observed, and $\langle v_{\perp}\rangle$ is measured and
presented in the middle panel, showing that the perpendicular expansion dominates when elongation has stopped, and
the mean speed of this expansion is 10 km~s$^{-1}$ initially, more than an order of magnitude slower than the elongation motion.
Later on when elongation has stopped, the ribbons still expand perpendicularly at a mean speed of 3 km~s$^{-1}$.
The mean electric field due to such an expansion is derived to be 5 V cm$^{-1}$ at about 16:40 UT, 10 minutes after the flare onset,
and decreases afterwards as the perpendicular expansion slows down.

Finally, it would be interesting to learn how flare loops are sheared with respect to the PIL. Unfortunately {\it TRACE} observations
of flare loops in this event are scarce, so we only measure the inclination of the conjugate ribbon brightenings
with respect to the photospheric PIL.  We assume the connectivity between newly brightened ribbon fronts, and measure the change 
of the inclination angle with respect to the PIL. Because of bi-directional spreading of the ribbons,
each ribbon has two newly brightened fronts. Ribbon observations alone do not provide the connectivity
between two pairs of the fronts. The inclination is therefore measured by assuming both connectivities, and the measurements
are presented in Figure~\ref{0513_2}c. For yet another ambiguity, the PIL in this region is curved rather than nearly straight.
In spite of these uncertainties, it is clear that this flare configuration possesses the largest shear among 
the three events. If we translate this inclination angle to the guide field relative to the outflow field, $B_g/B_o \approx 5$ initially, and then gradually
decreases to unity.

A filament is clearly visible along the PIL \citep{Yurchyshyn2006}. It disappears during the flare, and shows up again hours after the flare.
There is ambiguity as to whether the filament is partially erupted \citep{Tripathi2009} or does not disrupt at all but just 
experiences thermal disappearance \citep{Liu2007}. Again, it is not clear
how filament activity would be related to the bi-directional spread of ribbons. 

Of the three flares, this event is the most energetic, and hard X-ray emission up to 100~keV is observed by {\it RHESSI}.
Superimposed in Figure~\ref{0513_2}c is the {\it RHESSI} observed hard X-ray data counts in 25 -- 50 keV, which
is most likely produced by non-thermal electron beams impacting the chromosphere \citep{Liu2013}. The peak of this
non-thermal emission occurs at 16:43 UT. HXR spectral analysis (not shown here) also suggests that, at this moment, the photon spectrum
is hardest. At this time, elongation of the ribbons has mostly stopped, or the ribbon has attained its maximum length, and the perpendicular expansion starts to slow down. The observation that HXR emission peaks after the elongation phase is consistent
with an earlier report by \citet{Qiu2010} on the Bastille-day event. From these few examples, it therefore appears
that significant non-thermal particle production usually occurs after the elongation phase.

\subsection{Other Cases}

The events studied above provide examples of different kinds of ribbon elongation, one showing persistent slow
anti-parallel elongation, one with fast parallel elongation, both residing in magnetic fields of
low to moderate shear, and the last event with fast bi-directional elongation in a strongly sheared configuration. In addition to these cases, we 
summarize results of ribbon motion in other events presented in previous studies. These include the
SOL2000-07-04 X5.7 flare (Bastille-day flare) analyzed in \citet[][ and references therein]{Qiu2010}, 
SOL2005-01-15 X1.5 flare analyzed by \citet{Liu2010, Cheng2012}, and SOL2004-11-07 X2.0 flare by \citet{Longcope2007, Qiu2009}. 
Apparent ribbon motions have been measured in all these flares. The first two flares each show two stages of 
energy release taking place at different locations along the PIL, and motion is measured separately in each stage. 
For the last flare, flare ribbon brightening starts at multiple locations and is tracked separately at these different places.
All these events were observed in the UV 1600 \AA\ bandpass by {\it TRACE}. The observation cadence of the SOL2000-07-04 X5.7 flare is about 30~s, that of the SOL2004-11-07 X2.0 
event and SOL2005-01-15 X1.5 event are 7~s and 2~s, respectively.

\begin{figure}
\epsscale{1.0}
\plotone{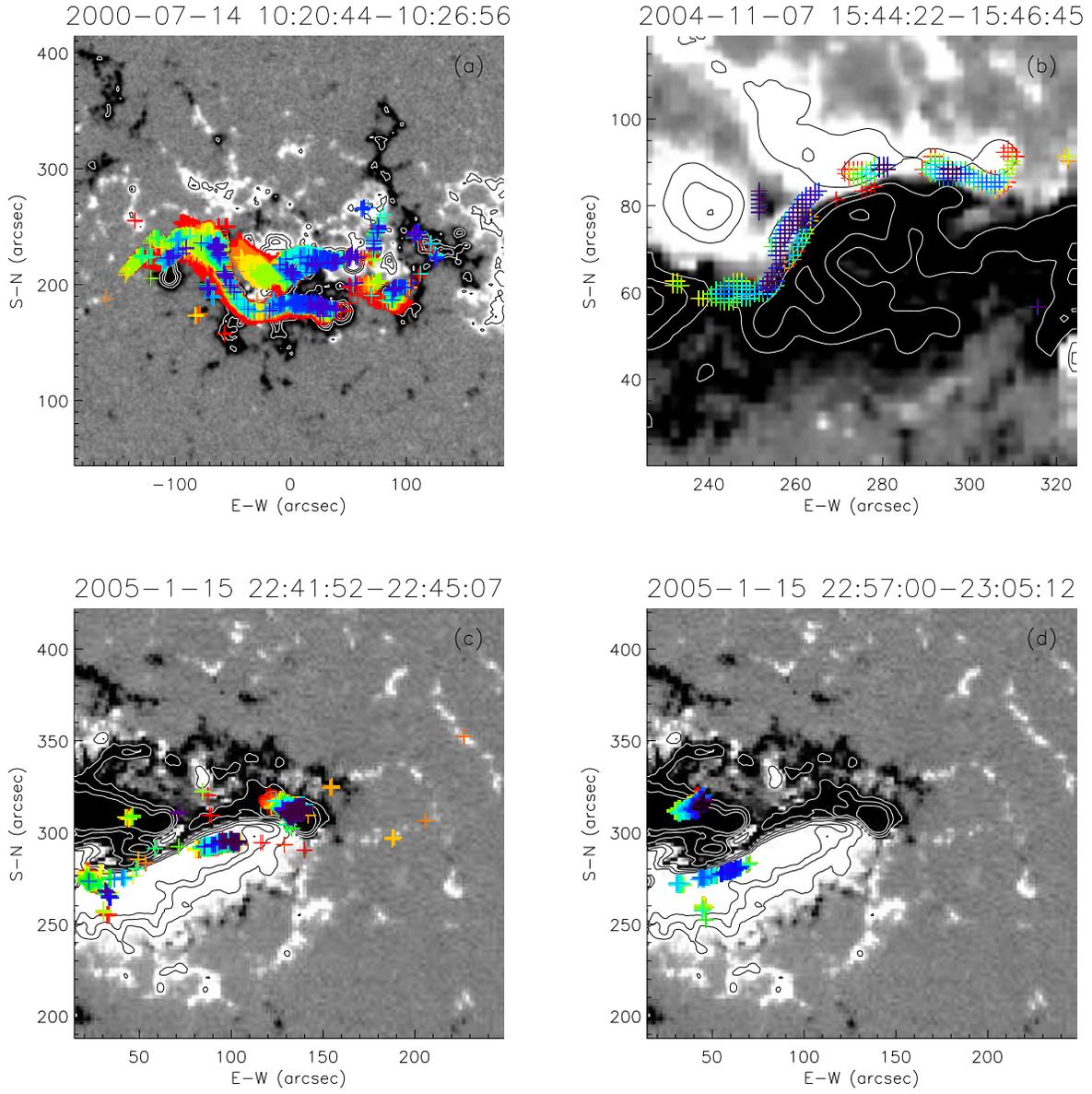}
\caption{Positions of newly brightened UV ribbon fronts in the first few minutes at the flare onset for 3 events. 
Color code from violet to blue, green, yellow, and orange indicates the time of the ribbon fronts. } \label{all}
\end{figure}

Figure~\ref{all} gives a general view of these three flares. For each flare, positions of the newly brightened ribbon fronts
are plotted on a longitudinal magnetogram by MDI, with color from violet to red indicating the time lapse. The definition of newly
brightened ribbon fronts depends on the threshold data counts. In this figure, pixels with counts larger than 10 times 
the quiescent background are plotted. To measure the ribbon motion, a few thresholds are used, and the position 
of the newly brightened ribbon fronts is the mean measurement with these thresholds. Note that in this paper, 
we only plot evolution of the ribbon fronts in the first few minutes to focus on the elongation, which
only occurs for a short time after the flare onset. Apparent motion of ribbon fronts, including the separation motion, 
over the entire flare duration, has been presented in previous studies.

The SOL2000-07-04 X5.7 flare (event 4 in Table 1) starts from the western end of the active region shortly after 10:00 UT, which is initiated by a 
filament eruption off that location. The observing cadence of $\ge$ 30~s does not allow us to track the motion of
brightening reliably at the start of the flare. From 10:20 UT, a second stage of energy release takes place,
and the flare spreads eastward toward the center of the active region, as indicated by the color map in Figure~\ref{all}a. 
It is noted that brightening starts at a couple of locations along each ribbon, and at some locations of strong emission, 
an organized pattern of elongation is observed. In particular, between 10:20:44 and 10:22:59 (violet to blue), eastward 
spreading is seen in both ribbons in the west of the AR. The mean speed of spreading is 65 km s$^{-1}$, which is more reliably  
measured in the negative ribbon, and the uncertainty is simply the standard deviation of the linear fit to derive the mean speed. 
Afterwards, eastward elongation is observed further toward the east of the AR, especially in 
the positive ribbon. From 10:24:10 to 10:26:56 (green to red), the mean speed of the spreading in the positive ribbon 
is 96 km s$^{-1}$. We note that a large section of the negative ribbon during this stage is brightened simultaneously with a less clear
pattern of motion, so we do not measure the speed there. \citet{Fletcher2001} have observed the same pattern of ribbon evolution.
From the magnetogram, the macroscopic electric current along and above the PIL 
runs westward\footnote{Note that \citet{Qiu2010} mistakenly gave the direction of the reconnection electric field as eastward.}.
Therefore, the overall trend of ribbon elongation is anti-parallel in this event.
The inclination angle of post-reconnection connectivity is estimated to be below 50$^{\rm o}$ at around 10:21 UT, and 50$\pm 10^{\rm o}$ from 10:25 -- 10:27 UT
\citep{Qiu2010}, yielding $B_g/B_o \approx 1.0-0.8$.

The SOL2004-11-07 X2.0 flare (event 5) starts at the PIL, with brightening occurring at a few places. For this flare, \citet{Qiu2009}
measured the motion of brightening in a few magnetic cells throughout the flare duration. Figure~\ref{all}b displays the ribbon brightening only 
in the first 2.5 minutes. It shows that in the positive polarity, brightening starts at three places and spreads along the PIL
from two locations. In the westmost patch (P5 in \citet{Qiu2009}), elongation is bi-directional at 77 km s$^{-1}$ westward
and 20 km s$^{-1}$ eastward. The middle patch (P3 in \citet{Qiu2009}) spreads mostly eastward at a mean speed of 77 km s$^{-1}$.
In the negative polarity, brightening starts in one patch, which spreads bi-directionally at a speed of 40 km s$^{-1}$ westward and 92 km s$^{-1}$ eastward.
The motion pattern in this event is therefore rather complex with different behavior at different locations. The macroscopic electric current
runs eastward in this flare. There is a large uncertainty in the measurement of the initial inclination because the negative ribbon may be connected
to different patches in the positive polarity. Nevertheless, this initial inclination angle can be estimated to be smaller than 20 degrees, yielding
$B_g/B_o$ greater than 3.

The SOL2005-01-15 X1.5 flare (event 6) also exhibits two stages of energy release at two different locations along the PIL \citep{Cheng2012}. The ribbon
brightening fronts are plotted in Figure~\ref{all}c and Figure~\ref{all}d, respectively, for these two stages. From 22:41:52 -- 22:45:07, 
brightening occurs in
the west of the AR with organized eastward elongation at average speeds of 49 and 35 km s$^{-1}$ in the positive and negative
ribbons, respectively. This flare was also observed by {\it RHESSI}. \citet{Cheng2012} found two HXR kernels at energies above 25 keV, which located
at the maximum UV emission along the two ribbons. The two HXR kernels also exhibit eastward elongation motion during the first 2 -- 3 minutes
at an average speed of 55 and 45 km s$^{-1}$ in the positive and negative fields, respectively \citep{Cheng2012}. Ten minutes later, a second
episode of energy release takes place in the east of the active region, and both ribbons appear to spread mostly eastward, at speeds
of 8 and 12 km s$^{-1}$, in the positive and negative fields, respectively. For this active region, the macroscopic electric
current runs eastward. Therefore, the overall elongation of the ribbons, at the initial stage of energy release, is parallel to the current
direction. \citet{Cheng2012} also estimated the inclination in these two stages, using both UV and HXR data; this angle is 40 and
75 degrees, respectively, at the start of each of the two stages of energy release.

\subsection{Properties of Ribbon Elongation and Magnetic Fields}

These observations, albeit from a rather small sample, demonstrate a variety of ribbon elongation patterns. Ribbons 
may elongate in a single direction, either parallel or anti-parallel to the current, or bi-directionally.  The parallel spreading occurs at a range of speeds
from a few kilometers per second to more than one hundred. These properties might be related to properties of the magnetic field or
electric current in the corona. The inclination angle of the post-reconnection connectivity is a frequently measured property. Apart from
this, we also estimate other properties, including the mean photospheric longitudinal magnetic field strength $\langle B_{ph} \rangle$
at the location of the initial brightening where elongation starts, the mean gradient of the photospheric magnetic field 
strength $\langle \nabla_{||} |B_{ph}|\rangle$ along the ribbon direction, and the mean total magnetic field strength 
in the corona $\langle B_{c}\rangle$ at the PIL and its gradient along the PIL. The coronal magnetic field is estimated by a potential field extrapolation 
at the approximate height of the flare loop top, estimated from the foot-point separation.
It is understood that all these measurements, especially the coronal magnetic field, carry large uncertainties. The uncertainties in $\langle B_{ph}\rangle$
and $\langle B_c\rangle$ are quoted as the standard deviation of the measured field in multiple pixels along the newly brightened ribbon or the coronal PIL, and 
the uncertainty in $\langle\nabla_{||} |B_{ph}|\rangle$ is simply the standard deviation of the linear fit to derive the gradient, which is the lower limit of real uncertainties. 

\begin{figure}
\epsscale{1.0}
\plotone{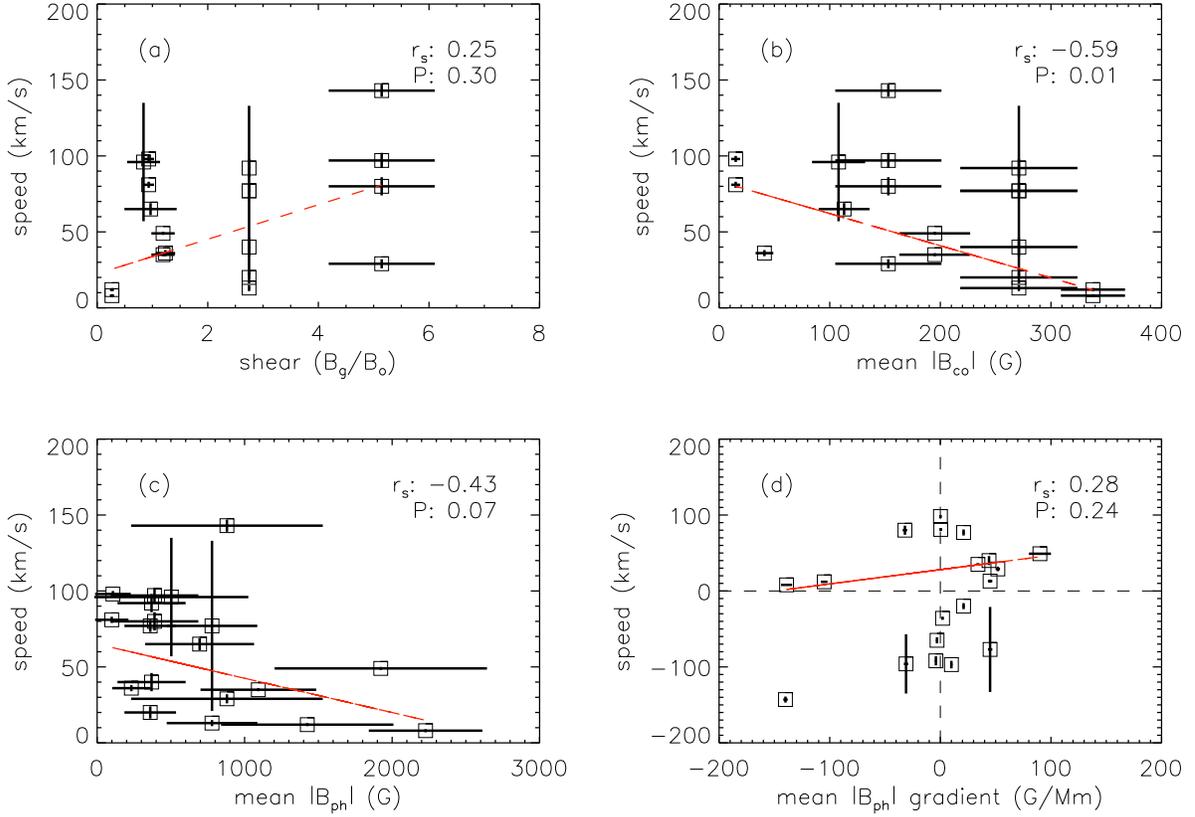}
\caption{Scatter plot of the ribbon elongation speed versus magnetic properties: (a) unsigned speed versus the magnetic shear
in terms of the ratio $B_g/B_o$; (b) unsigned speed versus the mean coronal magnetic field strength; (c) unsigned speed versus the mean photospheric magnetic field
strength; (d) signed speed with respect to the local magnetic gradient along the PIL (see text). In each panel, the red dashed
line shows the linear fit to the data, and $r_s$ and $P$ give the Spearsman rank correlation coefficient and the significance 
of its deviation from zero, respectively.} \label{plots}
\end{figure}

These measurements for the 6 events, including different episodes of energy release for some events, are listed in Table 1,
and comparison of their properties is given in Figure~\ref{plots}, showing the unsigned or signed ribbon elongation
speed versus various magnetic properties estimated above.

From this small sample categorized into three groups (parallel, anti-parallel, or bi-directional elongation), it first appears that the two events
(3 and 5) exhibiting bi-directional elongation have the strongest shear.  This would indicate a large relative guide field 
in a 2.5d approximation with $B_g/B_o$ greater than 3 in the first few minutes. For the other four events, with two exhibiting parallel
elongation and the other two anti-parallel elongation, there is no clear distinction in the shear: all of them have a weak to
moderate shear, with $B_g/B_o$ ranging between 0.3 and 1.2. In terms of elongation speed, there is a vague trend that
events (or locations) with greater shear move faster. The strong shear events (3 and 5), though exhibiting a variety
of elongation speeds at different locations, have the maximal speeds close to 100 km s$^{-1}$, whereas the slowest motion 
is measured in the second stage of event 6 (8 -- 12 km s$^{-1}$), which is associated with the weakest shear ($\theta_f = 75^{\rm o}$, and $B_g/B_o = 0.3$).
However, for the entire sample, the correlation between the speed and shear is low (Figure~\ref{plots}a).

Another possible factor in governing the elongation speed is the coronal magnetic field strength itself. This comparison
is given in Figure~\ref{plots}b, showing that the ribbon speed is roughly anti-correlated with the total magnetic field strength at
the approximate height of the loop top. This trend is partly governed by the two extreme cases, the second stage of event 6 with the slowest motion in strongest
magnetic fields, and event 2 with the lowest mean magnetic field but rather high speed close to 100~km s$^{-1}$. 
If at all, this is so far the strongest trend in all these comparisons. It is not clear why
the elongation speed is nearly anti-correlated with the coronal magnetic field.

If we consider the guide field strength to be the product of the coronal magnetic field strength and the cosine of the inclination angle, the correlation
in Figure~\ref{plots}a and anti-correlation in Figure~\ref{plots}b lead to a scatter between the measured speed and estimated
guide field strength. Therefore, the present observations, which carry a large uncertainty in our estimate of $B_g$, do not support 
the scenario that elongation is solely governed by Alfv\'en waves. Furthermore, the
maximum measured speed is nearly an order of magnitude smaller than the characteristic coronal Alfv\'en speed at the coronal magnetic field
strength presented in Table 1.

Since the coronal magnetic field in the present study is estimated very roughly from the potential field extrapolation, we also
examine the mean photospheric magnetic field at the locations of initial brightenings. 
Note that only the longitudinal magnetic field strength is measured, because four events are observed
by MDI, which does not have vector field measurements. Figure~\ref{plots}c shows that the speed appears
to anti-correlate with $\langle |B_{ph}|\rangle$, though to less of an extent than in Figure~\ref{plots}b. 

It is noted that motion of the conjugate ribbon fronts in opposite magnetic fields is mostly asymmetric. Since the same amount of positive and negative
magnetic flux participates in reconnection, ribbons should move faster in weaker magnetic fields; such a flux balance rule, namely $v_+B_+ \approx v_-B_-$
($+$ and $-$ indicating properties in the positive and negative magnetic fields, respectively),
is observed within a factor of two for these 6 events, if we ignore the perpendicular motion.
This fact partly contributes to the anti-correlation between $v_{||}$ and $\langle |B_{ph}|\rangle$ shown 
in Figure~\ref{plots}c, but cannot explain the stronger anti-correlation of the elongation speed with the 
coronal magnetic field. It is also noted that bi-directional elongation (in events 3 and 5) from one location is usually not symmetric.
To understand whether these asymmetries are due to the {\em local} magnetic field distribution, Figure~\ref{plots}d compares the 
signed speed with the gradient of the local magnetic field strength along the PIL. A plausible hypothesis could be that ribbon spreading tends to 
proceed in the direction the magnetic field decreases. Therefore, in Figure~\ref{plots}d, the signed speed is plotted with respect to the signed 
magnetic gradient, with the positive sign indicating ribbon elongation or magnetic field {\em decrease} in the direction of (i.e., parallel to) 
the electric current, and negative sign indicating ribbon elongation or magnetic field strength {\em decrease} in the direction opposite to 
(or anti-parallel to) the electric current. Figure~\ref{plots}d, although exhibiting a weak correlation between the two, shows very large
scatter and low significance. In this sample, the strongest supporting case is the positive ribbon in event 3, which spreads at high speed
away from the sunspot with quickly decreasing magnetic field; however, in this same event, negative ribbons appear to spread along the 
direction where magnetic field increases. The second stage of event 6 shows a similar counter-example, with ribbon spreading in the 
direction along which the magnetic field {\em increases} quickly. Event 2 shows an extreme case of fast spreading of ribbons in a pair of 
plages with a mean magnetic gradient close to zero. The rather random distribution of the elongation speed with respect to $\langle \nabla_{||} |B_{ph}|\rangle$
suggests that the local magnetic gradient is not a governing factor for ribbon elongation speed.

The mean gradient of the coronal magnetic field along the PIL is also estimated. The coronal field from extrapolation exhibits a much smoother 
distribution along the PIL but a similar trend of gradients to that in the photosphere, although the magnitude of the gradient is smaller 
by one to two orders of magnitude (not shown). 

Furthermore, the gradient of the photospheric magnetic field strength in the direction perpendicular to the PIL is estimated at locations of initial brightenings (not shown). It is observed that ribbon brightening tends to start
from the edge of a plage or penumbra close to the PIL; therefore, 
in these regions of initial brightening, the unsigned magnetic field strength increases away from the PIL. The magnitude of this perpendicular
gradient ranges from a few tens to a few hundreds gauss per Mm, about an order of magnitude greater than the parallel gradient in
the photosphere. However, it is less clear how to relate the locally measured perpendicular gradient of the photospheric magnetic 
field with the coronal configuration around the reconnection region, even in a 2.5d approximation. These measurements also 
do not illustrate any pattern of correlation or anti-correlation with ribbon elongation in this sample.

\citet{Liu2010} also measured the H$\alpha$ and hard X-ray foot-point motion
in Event 6, and found that H$\alpha$ and hard X-ray kernels spread in the direction along which the overlying magnetic field
decays with height more quickly. Our analysis with this small sample, however, does not reveal a correlation between the 
direction of ribbon spreading and the decay index of the overlying field along the PIL.

%
%

In addition, Table 1 records the mean speed of the perpendicular expansion $v_{\perp}$ of the entire ribbon when it is fully formed.
This average speed ranges from a few to up to 10 km s$^{-1}$, generally smaller than $v_{||}$, and we do not find a correlation between the elongation speed $v_{||}$ and the
perpendicular speed $v_{\perp}$.

In summary, for the given small sample of six two-ribbon flares, the ribbon elongation speed
appears to be anti-correlated with the photospheric magnetic field strength as well as the mean coronal magnetic field 
along the PIL. On the other hand, we do not find that magnetic properties or motion speeds associated 
with parallel elongation are substantially different from those of anti-parallel elongation.

\section{Summary and Discussion}
We present an analysis of apparent motion of flare ribbon brightenings along the magnetic polarity inversion line, the so-called
``elongation" motion, in the early phase of three flares observed in the UV 1600 \AA\ bandpass. They each exhibit a 
different pattern of ribbon elongation. In one event, ribbon brightening spreads along the PIL in the direction opposite to the macroscopic electric current in the corona (anti-parallel elongation) for as long as one hour 
at a mean speed of 11 km s$^{-1}$. Another event exhibits two ribbons spreading quickly in the same
direction as the current (parallel elongation) at a mean speed of nearly 100 km s$^{-1}$ for about 10 minutes. 
In the third event, brightening spreads quickly and bi-directionally. The first two flares were also 
observed in the EUV bandpasses by {\it AIA}, showing ordered spreading of flare EUV loops along the polarity inversion line
after the ribbon brightening, at mean speeds systematically smaller
than the ribbon spreading. These observations confirm the ``zipper" pattern of elongation motion in both
flare ribbons and flare loops produced by successive reconnection energy release
along the PIL. 

We measure the inclination angle $\theta$ of the line connecting conjugate foot-points with 
respect to the polarity inversion line of longitudinal magnetic field of the photosphere. With flare loops observed in the EUV 
171 \AA\ passband by {\it AIA}, this angle is also measured as the inclination of the observed loop tops with respect to the 
polarity inversion line. The two measurements are consistent with each other, both 
showing gradually decreased shear as the flare evolves and both ribbons and loops spread along the polarity inversion line.
The measured flare shear is also larger (i.e., smaller $\theta$) than that of the potential field, suggesting that 
flare loops usually are not yet relaxed to a potential configuration. With a 2.5d approximation of the flare arcade configuration, we estimate
the magnetic guide field $B_g$, the component along the current (and PIL) that does not participate in reconnection,
relative to the reconnection outflow field $B_o$ using the relation 
$B_g/B_o \approx {\rm cot}\theta$. We find that the event with bi-directional elongation has a strong shear
with $B_g/B_o \approx 5$, and the other events with uni-directional elongation have a moderate
shear with $B_g/B_o \approx 0.4 - 1.2$.

We review properties of elongation in another 3 X-class flares analyzed in previous studies, and compare 
these properties with magnetic properties in flare regions for all six events. It is observed that, in this
small sample, ribbon elongation speed is greater in events with weaker mean coronal or photospheric
magnetic field, but is less correlated with the inclination angle $\theta$ or the magnetic gradient $\langle\nabla_{||}|B|\rangle$ 
along the PIL measured either locally in the photosphere or in the corona.

These results demonstrate the difficulty to identify physical mechanisms governing the ribbon elongation motion. 
As much as we have attempted to infer reconnection properties from signatures in the lower atmosphere, where 
the magnetic field can be measured, there are large uncertainties in deriving the magnetic field in the corona. 
The difficulties are even greater in inferring properties within the reconnecting current sheet. The two events 
of bi-directional elongation exhibit a strong shear with $B_g/B_o > 3$, which might support the scenario of Alfv\'en waves.  
On the other hand, in these events, the motion speed of $\le 100$ km s$^{-1}$ is an order of magnitude smaller
than the coronal Alfv\'en speed for the given mean coronal magnetic field in Table 1 
(if $n \sim 10^9$ cm$^{-3}$). It is also shown that, for the entire sample, the present measurements 
(with very large uncertainties in the coronal magnetic field) cannot establish the
evidence that the elongation speed $v_{||}$ scales with the guide field strength $B_g$.
Alternatively, because of the large uncertainty in coronal magnetic field measurements, we may 
estimate the guide field strength based on the ribbon elongation speed. This yields $B_g \approx 2$ G for the 
measured fastest speed at about 100 km s$^{-1}$. At the strongest shear $\theta \approx 10 - 20^{\rm o}$, 
we find the outflow magnetic field component to be less than 1 G, significantly smaller than the coronal field from extrapolation. 

The other population of events of parallel or anti-parallel elongation both exhibit a range of 
initial motion speed from a few to nearly a hundred kilometers per second with $B_g/B_o \approx 
0.3 - 1.2$. If we consider these motions as being produced by drifting of current-carrying
particles, the inferred thickness of the reconnection current sheet is of order 10$^{5-6}$ cm
(again if $n \sim 10^9$ cm$^{-3}$). It is interesting to see that, whereas this scale is larger, by 2 -- 3 orders of magnitude, than some 
theoretical estimates (such as in the Sweet-Parker and Hall reconnection models), it is smaller by 1 -- 2 orders 
of magnitude than the reported thickness of long current sheets trailing behind CMEs \citep[][and references
therein]{Lin2015}. The thickness of a flare current sheet remains a matter of debate \citep[e.g., ][]{Lin2015}.

Results from this small sample also show an anti-correlation between elongation speed and the magnetic field strength.
The anti-correlation shown for the photospheric magnetic field is partly attributed to the reconnection
flux balance between positive and negative fields. Yet the anti-correlation with the coronal
magnetic field strength is not understood. Energy release in strong magnetic fields tends to be greater, which might in general require a fast reconnection
or a large characteristic speed in the system. The mean magnetic field strength in this sample varies from one
event to another by one and a half orders of magnitude; the total reconnection flux as well as the peak reconnection
rate (in units of Maxwell per second) in these events also vary by one and a half orders of magnitude, with
flares in stronger magnetic fields indeed exhibiting greater reconnected flux and reconnection rate. On the other hand,
measurements in this paper show that flares in stronger magnetic field do not spread faster.

All the events in this sample are accompanied by coronal mass ejections, but few by erupting filaments. 
The SOL2000-07-14 flare starts at the location where a filament lifts off, and for the SOL2005-01-15 event,
\citet{Liu2010} suggested that flare kernels spread in the direction along which the overlying magnetic field decreases with height
more rapidly. But for most of these events, we do not find a clear association between the location or
elongation speed of initial ribbon brightening and filament or CME dynamics. 
On the other hand, \citet{Hu2014, Priest2017} have discussed how the geometry and sequence of 
reconnection as reflected in ribbon motion may lead to formation of flux ropes. 

To achieve an understanding of possible governing mechanisms, it will be helpful to apply the analysis
to a larger sample of ribbon elongation in both eruptive flares and compact flares \citep[e.g. ][]{Veronig2015}
with simple configurations. Since a more accurate inference of coronal magnetic fields cannot be achieved
at present, we hope that analysis of a larger sample will help clarify whether there are 
trends relating ribbon motion direction and speed with the shear that allows us to test proposed mechanisms.

We also acknowledge that the magnetic configuration at or near the reconnection site is quite different
from a potential field. For instance, reconnection may occur at the top of a cusp structure, at an altitude
possibly much greater than the top of flare loops, which have retracted from the reconnection site.
In such a circumstance, the field strength may be much smaller than estimated in this paper, and the
characteristic speed of reconnection spread could be smaller as well. To verify this scenario,
it is useful to explore suitable and simultaneous coronal observations from a different 
vantage point, such as those by {\it STEREO}, which may allow us to observe the cusp structure and estimate the height
and magnetic field there.

\acknowledgments Authors thank the referee for careful review and constructive comments that help improve the paper.
JQ, DWL, and PAC gratefully acknowledge support by NSF SHINE collaborative grant AGS-1460059. SDO is a mission of NASA's Living With a Star Program.

\bibliography{asym}

\newpage

\begin{deluxetable}{lllllllllll}
\tabletypesize{\scriptsize}
\rotate \tablecolumns{10} \tablewidth{0pt} \tablecaption{Properties of Flare Ribbon Elongation\label{flareinfo}}
    \tablehead{
    \colhead{ } &
    \colhead{date, magnitude$^{a,b}$} &
    \colhead{position} &
    \colhead{time$^c$} &
    \colhead{direction$^d$} &
    \colhead{$\langle v_{||}\rangle^e$ } &
    \colhead{$\langle v_{\perp} \rangle^f$ } &
    \colhead{$\theta_f^g$} &
    \colhead{$B_g/B_o$} &
    \colhead{$\langle B_{ph} \rangle^h$, $\langle B_c \rangle^i$} &
    \colhead{$-\langle\nabla_{||} |B_{ph}| \rangle^j$ }}
\startdata
1 & 2011-09-13 C2   & AR11289 N22W14  & 22:51 & anti-parallel (P) & -36$\pm$2 		& 0 & 39$\pm$4  & 0.7  & 232$\pm$130, 41$\pm$8     & 1.7$\pm$0.1 \\
2 & 2011-12-26 C3.7 & AR11384 N13W14  & 11:18 & parallel (P)      &  +81$\pm$2 		& 2 & 47$\pm$3  & 0.9  & 99$\pm$113, 15$\pm$4      &  0.4$\pm$0.2  \\ 
  &                 &                 &       & parallel (N)      &  +98$\pm$2 		& 2 &          &      & -107$\pm$121, 15$\pm$4    & 0.1$\pm$0.4  		\\
3 & 2005-05-13 M8.0 & AR10759 N12E05  & 16:34 & bi-directional (P) & +29$\pm$3,-143$\pm$4& 8 & 11$\pm$2 & 5.1  & 880$\pm$650, 153$\pm$48   & 52$\pm$2, -140$\pm$2  \\ 
  &                 &                 &       & bi-directional (N) & +80$\pm$6,-97$\pm$5 &14 &	    &  	   & -389$\pm$297, 153$\pm$48  & -32$\pm$2, 10$\pm$1  \\ 
4 & 2000-07-14 X5.7 & AR9077 N17E01   & 10:21 & anti-parallel (N) & -65$\pm$4 		& 8 & 46$\pm$14 & 1.0  & -695$\pm$369, 113$\pm$23  & -3.0$\pm$0.2 \\
  &                 &                 & 10:24 & anti-parallel (P) & -96$\pm$39 		& 8 & 50$\pm$10 & 0.8  & 503$\pm$522, 108$\pm$24   & -31$\pm$2 \\
5 & 2004-11-07 X2.0 & AR10696 N09W22  & 15:44 & bi-directional (P3)& +13$\pm$2,-77$\pm$56& 10 & $<$20    & $>$3 & 779$\pm$307, 271$\pm$53   & 45$\pm$2  	\\
  &                 &                 &       & bi-directional (P5)& +77$\pm$4,-20$\pm$4 & 10 &      &      & 360$\pm$175, 271$\pm$53   & 21$\pm$0.5  		 \\
  &                 &                 &       & bi-directional (N) & +40$\pm$6,-92$\pm$  & 2 &        &      & -369$\pm$231, 271$\pm$53  & 44$\pm$1, -4$\pm$0.5 \\
6 & 2005-01-15 X2.6 & AR10720 N13W04  & 22:42 & parallel (P)      & +49$\pm$1 (+55$^k$) & 9 & 40$\pm$5  & 1.2  & 1923$\pm$720, 195$\pm$32  & 96$\pm$10, -85$\pm$5 \\
  &                 &                 &       & parallel (N)      & +35$\pm$1 (+45$^k$) & 22&          &      & -1093$\pm$392, 195$\pm$32 & 34$\pm$1.5   			 \\
  &                 &                 & 22:57 & parallel (P)      & +8$\pm$1 		& - & 75$\pm$2  & 0.3  & 2227$\pm$385, 338$\pm$29  & -139$\pm$5  			 \\
  &                 &                 &       & parallel (N)      & +12$\pm$1 		& - &	    &      & -1425$\pm$584, 338$\pm$29 & -105$\pm$3 			\\

\enddata
\tablenotetext{a}{References for these events are as below: 
(1) \citet{Hu2014, Qiu2016}; 
(2) \citet{Cheng2016}; 
(3) \citet[][and references therein]{Liu2013}; 
(4) \citet[][and references therein]{Qiu2010}; 
(5) \citet{Longcope2007, Qiu2009};  
(6) \citet{Liu2010, Cheng2012}.}
\tablenotetext{b}{Magnitude is based on GOES classification.}
\tablenotetext{c}{Time refers to the start time when the elongation motion is measured.}
\tablenotetext{d}{Direction of elongation is given with respect to the flow direction of the electric current $\vec{J}$.
For each event, speed is reliably measured in either the positive fields (P), or negative fields (N), or both.
In some events, it is measured in multiple pieces of ribbons in the same polarity (such as P3 and P5 in event No. 5).}
\tablenotetext{e}{Elongation speed in units of kilometers per second. Positive speed ("+") refers 
to the speed in the direction of the current $\vec{J}$, and negative speed ("-") refers
to the speed in the direction opposite to the current directiion.}
\tablenotetext{f}{Mean perpendicular speed (in kilometer per second) of the entire ribbon measured when the ribbon is well formed.}
\tablenotetext{g}{Shear angle (degrees) measured from conjugate foot-points at the initial brightenings, except in event 1, where
the inclination angle measured from the first visible flare loops is used.}
\tablenotetext{h}{Mean photospheric longitudinal magnetic field (in Gauss) at the ribbon locations.}
\tablenotetext{i}{Mean strength of total coronal magnetic field (in Gauss) from potential field extrapolation, along the 
PIL at the approximate height of the flare loop top. For events with energy release at different locations during
different stages (e.g., events 4 and 6), the coronal field is estimated at different locations along the PIL.}
\tablenotetext{j}{Gradient of longitudinal magnetic field strength (Gauss per Mm) along the newly brightened sections of ribbons, along
which the elongation speed is measured in the first few minutes of the flare. The sign of the gradient is given with respect to the direction of the
electric current. Positive gradient indicates that the magnetic field {\em decreases} along the direction of $\vec{J}$, 
whereas negative gradient refers to {\em decrease} of magnetic field in the opposite direction of $\vec{J}$.}
\tablenotetext{k}{These are speeds of HXR foot-points \citep{Cheng2012}.} 
\end{deluxetable}

\newpage

\newpage

\end{document}